\documentclass[twocolumn,twocolappendix]{aastex631}
\usepackage{color}

\newcommand{\paper}{paper}
\newcommand{\p}{\phantom{$-$}}
\newcommand{\z}{\phantom{0}}
\newcommand{\m}{$-$}
\newcommand{\cifull}{\mbox{\rm [\ion{C}{1}] $^3P_1\text{--}^3P_0$}}
\newcommand{\ci}{\mbox{\rm [\ion{C}{1}]($1\text{--}0$)}}
\newcommand{\cishort}{\mbox{\rm [\ion{C}{1}]}}
\newcommand{\siii}{\mbox{\rm [\ion{S}{3}]}}
\newcommand{\sii}{\mbox{\rm [\ion{S}{2}]}}

\newcommand{\htcn}{\mbox{\rm H$^{13}$CN($1\text{--}0$)}}
\newcommand{\cch}{\mbox{\rm C$_2$H($1\text{--}0$)}}
\newcommand{\hcn}{\mbox{\rm HCN($1\text{--}0$)}}
\newcommand{\hcop}{\mbox{\rm HCO$^+$($1\text{--}0$)}}
\newcommand{\hnc}{\mbox{\rm HNC($1\text{--}0$)}}
\newcommand{\hcccn}{\mbox{\rm HC$_3$N}}
\newcommand{\hcccnten}{\mbox{\rm HC$_3$N($10\text{--}9$)}}
\newcommand{\nthp}{\mbox{\rm N$_2$H$^+$($1\text{--}0$)}}
\newcommand{\methanol}{\mbox{\rm CH$_3$OH($2_K\text{--}1_K$)}}
\newcommand{\cs}{\mbox{\rm CS($2\text{--}1$)}}
\newcommand{\hcccneleven}{\mbox{\rm HC$_3$N($11\text{--}10$)}}
\newcommand{\hcccntwelve}{\mbox{\rm HC$_3$N($12\text{--}11$)}}
\newcommand{\ceoone}{\mbox{\rm C$^{18}$O($1\text{--}0$)}}
\newcommand{\tcoone}{\mbox{\rm $^{13}$CO($1\text{--}0$)}}
\newcommand{\cnl}{\mbox{\rm CN($1_{1/2}\text{--}0_{1/2}$)}}
\newcommand{\cnh}{\mbox{\rm CN($1_{3/2}\text{--}0_{1/2}$)}}
\newcommand{\coone}{\mbox{\rm CO($1\text{--}0$)}}
\newcommand{\kms}{\mbox{km s$^{-1}$}}

\newcommand{\MPIA}{\affil{Max-Planck-Institut f\"{u}r Astronomie, K\"{o}nigstuhl 17, D-69117, Heidelberg, Germany}}
\newcommand{\Nichidai}{\affil{Department of Physics, General Studies, College of Engineering, Nihon University, 1 Nakagawara, Tokusada, Tamuramachi, Koriyama, Fukushima, 963-8642, Japan}}
\newcommand{\NAOJ}{\affil{National Astronomical Observatory of Japan, 2-21-1 Osawa, Mitaka, Tokyo, 181-8588, Japan}}
\newcommand{\IoA}{\affil{Institute of Astronomy, School of Science, The University of Tokyo, 2-21-1, Osawa, Mitaka, Tokyo 181-0015, Japan}}
\newcommand{\UFukushima}{\affil{Faculty of Symbiotic Systems Science, Fukushima University, Fukushima 960-1296, Japan}}

\newcommand{\SOKENDAI}{\affil{Department of Astronomy, School of Science, The Graduate University for Advanced Studies (SOKENDAI), 2-21-1 Osawa, Mitaka, Tokyo, 181-1855 Japan}}
\newcommand{\ISEE}{\affil{Institute for Space-Earth Environmental Research, Nagoya University, Furo-cho, Chikusa-ku, Nagoya, Aichi 464-8601, Japan}}
\newcommand{\UNagoya}{\affil{Division of Particle and Astrophysical Science, Graduate School of Science, Nagoya University, Furocho, Chikusa-ku, Nagoya, Aichi 464-8602, Japan}}
\newcommand{\UJoetsu}{\affil{Joetsu University of Education, Yamayashiki-machi, Joetsu, Niigata 943-8512, Japan}}
\newcommand{\RESCEU}{\affil{Research Center for the Early Universe, School of Science, The University of Tokyo, 7-3-1 Hongo, Bunkyo-ku, Tokyo 113-0033, Japan}}
\newcommand{\MPE}{\affil{Max-Planck-Institut f\"ur Extraterrestrische Physik (MPE), Giessenbachstr. 1, D-85748 Garching, Germany}}

\received{May 12, 2022}
\revised{June 26, 2021}
\accepted{July 13, 2022}

\shorttitle{PCA for ALMA Spectral Scan Data for NGC~1068}
\shortauthors{T. Saito et al.}

\graphicspath{{./}{figures/}}

\begin{document}
\title{AGN-driven Cold Gas Outflow of NGC~1068 Characterized by Dissociation-Sensitive Molecules}

\correspondingauthor{Toshiki Saito}
\email{toshiki.saito@nao.ac.jp}

\author[0000-0002-2501-9328]{Toshiki Saito}\NAOJ\Nichidai
\author[0000-0001-6788-7230]{Shuro Takano}\Nichidai
\author[0000-0002-6824-6627]{Nanase~Harada}\NAOJ\SOKENDAI
\author[0000-0002-8467-5691]{Taku~Nakajima}\ISEE
%\author{author list TBD}
\author[0000-0002-3933-7677]{Eva~Schinnerer}\MPIA
\author[0000-0001-9773-7479]{Daizhong~Liu}\MPE
\author[0000-0002-9695-6183]{Akio~Taniguchi}\UNagoya
\author[0000-0001-9452-0813]{Takuma~Izumi}\NAOJ\SOKENDAI
\author{Yumi~Watanabe}\UFukushima
\author[0000-0001-9720-8817]{Kazuharu~Bamba}\UFukushima
\author[0000-0002-4052-2394]{Kotaro~Kohno}\IoA\RESCEU
\author[0000-0003-0563-067X]{Yuri~Nishimura}\IoA\NAOJ
\author[0000-0002-9333-387X]{Sophia~Stuber}\MPIA
\author[0000-0001-9016-2641]{Tomoka~Tosaki}\UJoetsu

%%%%%%%%%%%%%%%%%%%%%%%%%%%%%%
%%%%%%%%%% Abstract %%%%%%%%%%
%%%%%%%%%%%%%%%%%%%%%%%%%%%%%%
%TC:ignore
\begin{abstract}
Recent developments in (sub-)millimeter facilities have drastically changed the amount of information obtained from extragalactic spectral scans. In this \paper, we present a feature extraction technique using principal component analysis (PCA) applied to arcsecond-resolution (1\farcs0--2\farcs0 $=$ 72--144~pc) spectral scan datasets for the nearby type-2 Seyfert galaxy, NGC~1068, using Band~3 of the Atacama Large Millimeter/submillimeter Array. We apply PCA to 16 well-detected molecular line intensity maps convolved to a common 150~pc resolution. In addition, we include the \siii/\sii\ line ratio and \cifull\ maps in the literature, both of whose distributions show remarkable resemblance with that of a kpc-scale biconical outflow from the central AGN. We identify two prominent features: (1) central concentration at the circumnuclear disk (CND) and (2) two peaks across the center that coincide with the biconical outflow peaks. The concentrated molecular lines in the CND are mostly high-dipole molecules (e.g., H$^{13}$CN, HC$_3$N, and HCN). Line emissions from molecules known to be enhanced in irradiated interstellar medium, CN, C$_2$H, and HNC, show similar concentrations and extended components along the bicone, suggesting that molecule dissociation is a dominant chemical effect of the cold molecular outflow of this galaxy. Although further investigation should be made, this scenario is consistent with the faintness or absence of the emission lines from CO isotopologues, CH$_3$OH, and N$_2$H$^+$, in the outflow, which are easily destroyed by dissociating photons and electrons.
\end{abstract}

%% https://astrothesaurus.org
%TC:endignore

%%%%%%%%%%%%%%%%%%%%%%%%%%%%%%
%%%%%%%% Introduction %%%%%%%%
%%%%%%%%%%%%%%%%%%%%%%%%%%%%%%
\section{Introduction} \label{sec:intro}
Outflows or jets driven by accreting supermassive black holes (active galactic nuclei; AGNs) are considered to play a crucial role in regulating star formation in massive galaxies, and thus, determine the evolution of galaxies in the universe \citep[e.g.,][]{Scannapieco04,ForsterSchreiber20}. In particular, outflows are considered to be capable of quenching star formation activities by heating and/or sweeping dense and cold molecular regions in the interstellar medium (ISM), where stars form \citep[negative AGN feedback; e.g.,][]{Sturm11,Chen22}. Thus, galactic cold (neutral atomic and molecular) outflows are important phenomena in galaxy evolution \citep{Veilleux20}. A recent observational study by \citet{Stuber21} suggested that a non-negligible number of star-forming galaxies (20\%--25\%) showed outflow-like features in carbon monoxide (CO) even at redshift $=$ 0.

With the establishment of the Atacama Large Millimeter/submillimeter Array (ALMA), astronomers have started to target lines fainter than the bright CO lines toward nearby galactic cold gas outflows, to inspect their physical and chemical properties \citep[e.g.,][]{Aalto12,Walter17,Aladro18,Harada18,Michiyama20a}. Concurrently, unbiased spectral line surveys have begun to show the molecular line richness and diversity in extragalactic systems \citep[e.g., ][]{Martin06,Rangwala11,Aladro15,Takano19,Martin21,denBrok22,Eibensteiner22} as well as in galaxies with known molecular outflow features \citep[e.g., ][]{Sakamoto21}.

With the increase in the number of detected molecular lines and the complex relation between the chemistry and physics behind each molecular line, it is becoming increasingly difficult to interpret line intensities and intensity ratios coherently. To address the increasing complexity, astronomers have utilized a commonly used unsupervised machine learning technique called principal component analysis (PCA; \citealt{Pearson01,Abdi10}), and succeeded in reducing the dimensionality, i.e., categorizing and interpreting molecular line diversity \citep[e.g.,][]{Costagliola11}.

Owing to the state-of-the-art high-sensitivity and high angular resolution (sub-)millimeter facilities provided by ALMA, we can currently study the spatial distribution of each molecular line at tens of pc in addition to its line intensity \citep[e.g.,][]{Takano14,Martin21}. The number of dimensions and features are continuing to increase. To quantify the morphology of molecular line intensity maps for galaxies, PCA has begun to be applied to the imaging spectral scan datasets of some nearby galaxies \cite[e.g.,][]{Meier05,Johnson18}.

In this \paper, we present an application of PCA to high-quality ALMA molecular line maps of the nearby type-2 Seyfert galaxy NGC~1068 ($\sim$13.97~Mpc; \citealt{Anand21}) captured as part of an imaging spectral scan campaign (\citealt{Takano14,Nakajima15,Tosaki17}; Nakajima et al. in preparation) and to multi-wavelength maps from the literature \citep{Mingozzi19,Saito22}. Subsequently, we describe the outflow features in the central kiloparsec (1-kpc) of NGC~1068 extracted by PCA and present a possible chemistry that coherently explains all PCA results. PCA provides qualitative constraints on the physics and chemistry happening in the center of NGC~1068 based on the morphology of each map. Thus, a quantitative analysis should be done to verify the results of PCA, although it is beyond the scope of this \paper.

%%%%%%%%%%%%%%%%%%%%%%%%%%%%%%
%%%%%%%%%%%% Data %%%%%%%%%%%%
%%%%%%%%%%%%%%%%%%%%%%%%%%%%%%
\section{Observations and Processing} \label{sec:data}
\subsection{ALMA observations, ancillary data, and data processing}
We conducted an imaging spectral scan at 3~mm toward NGC~1068 using Band~3 of ALMA with its 12-m array (2013.1.00279.S: PI $=$ T. Nakajima). The details of the observations can be found in Nakajima et al. (in preparation). We also downloaded and processed all archival Band~3 data covering NGC~1068 being publicly available as of summer 2021. We focus on data with a spatial resolution of 150~pc or better ($<$2\farcs08), which left eight projects (2011.0.00061.S, 2012.1.00657.S, 2013.1.00060.S, 2015.1.00960.S, 2017.1.00586.S, 2018.1.01506.S, 2018.1.01684.S, and 2019.1.00130.S), in addition to 2013.1.00279.S. Some projects have employed observations of the 7-m array and total power array of ALMA. However, for consistency we only used 12-m data, to match the maximum recoverable scale among the detected lines. In addition to the Band~3 data, we used the Band~8 \cifull\ map (2017.1.00586.S; hereafter \cishort\ map), which traces the cold neutral gas outflow of this galaxy \citep{Saito22}.

\begin{deluxetable*}{lccccccc}
%\tablenum{1}
\tablecaption{Line properties and imaging properties\label{tab:data}}
\tablewidth{0pt}
\tablehead{
\colhead{Line} & \colhead{$\nu_{\rm rest}$} & \colhead{$\theta_{\rm maj}\times\theta_{\rm min}$} & \colhead{$V_{\rm ch}$} & \colhead{$\sigma_{\rm ch}$} & \colhead{Med $\sigma_{\rm hex}$} & \colhead{SNR$_{\rm hex}$} & \colhead{$N_{\rm hex}$} \\
\colhead{} & \colhead{(GHz)} & \colhead{(\arcsec)} & \colhead{(\kms)} & \colhead{(K)} & \colhead{(K km s$^{-1}$)} & \colhead{(16$^{th}$--50$^{th}$--84$^{th}$--max)} & \colhead{} \\
\colhead{(1)} & \colhead{(2)} & \colhead{(3)} & \colhead{(4)} & \colhead{(5)} & \colhead{(6)} & \colhead{(7)} & \colhead{(8)}
}
%\decimalcolnumbers
\startdata
\htcn\        & \z86.339921 & 0.59$\times$0.52 & 13.58  & 0.112 & 0.89 & 5.1--14.5--21.9--35.3 & 21 \\
\cch\         & \z87.316898 & 0.56$\times$0.49 & \z6.71 & 0.146 & 0.70 & 5.3--7.0--12.4--18.5 & 64 \\
\hcn\         & \z88.631602 & 0.59$\times$0.52 & \z6.61 & 0.132 & 0.94 & 6.3--10.6--26.3--145.6 & 77 \\
\hcop\        & \z89.188525 & 0.56$\times$0.49 & 19.65  & 0.062 & 1.02 & 5.4--8.1--16.2--66.9 & 75 \\
\hnc\         & \z90.663568 & 2.03$\times$1.74 & \z9.70 & 0.010 & 0.28 & 7.5--14.1--33.2--165.1 & 77 \\
\hcccnten\    & \z90.979023 & 2.07$\times$1.76 & 19.33  & 0.006 & 0.15 & 5.8--15.3--34.1--57.5 & 39 \\
\nthp\        & \z93.173977 & 1.98$\times$1.71 & 18.87  & 0.006 & 0.17 & 4.9--10.2--21.4--41.9 & 41 \\
\methanol\    & \z96.744550 & 0.89$\times$0.78 & \z9.85 & 0.021 & 0.15 & 5.3--8.0--16.8--26.8 & 48 \\
\cs\          & \z97.980953 & 0.82$\times$0.72 & \z9.72 & 0.032 & 0.31 & 5.8--8.0--19.4--86.2 & 76 \\
\hcccneleven\ & 100.076385  & 0.57$\times$0.52 & 17.58  & 0.045 & 0.32 & 5.9--13.9--21.0--34.0 & 25 \\
\hcccntwelve\ & 109.173638  & 0.79$\times$0.70 & 19.47  & 0.017 & 0.24 & 6.2--16.3--28.6--45.5 & 24 \\
\ceoone\      & 109.782173  & 0.82$\times$0.72 & \z9.35 & 0.021 & 0.19 & 4.7--6.9--10.2--17.2 & 77 \\
\tcoone\      & 110.201354  & 0.68$\times$0.61 & \z9.98 & 0.034 & 0.50 & 8.2--11.3--17.7--31.8 & 77 \\
\cnl\         & 113.191279  & 0.43$\times$0.38 & \z7.77 & 0.088 & 0.64 & 7.2--11.1--30.3--54.9 & 55 \\
\cnh\         & 113.490970  & 0.43$\times$0.38 & \z7.75 & 0.168 & 1.94 & 5.5--8.9--19.4--50.3 & 70 \\
\coone\       & 115.271202  & 0.40$\times$0.35 & \z0.64 & 0.801 & 3.30 & 18.4--29.5--48.0--77.6 & 77 \\
\ci\          & 492.160651  & 0.70$\times$0.56 & \z2.38 & 0.062 & 1.29 & 12.6--27.8--89.8--199.5 & 64 \\
\enddata
\tablecomments{
Column 1: Line name.
Column 2: Rest frequency of line \citep{Endres16}.
Column 3: Major and minor axes (full width at half maximum) of original synthesized beam before convolution to 150~pc ($\sim$2\farcs08).
Column 4: Velocity resolution.
Column 5: Noise root mean square (RMS) at velocity resolution and original synthesized beam size. RMS is measured at each channel, and median value of RMS histogram is listed. This could be different from value listed in Nakajima et al. (in preparation), which summarizes 2013.1.00279.S, because archival ALMA data are added to 2013.1.00279.S, to increase sensitivity of each line.
Column 6: Median noise RMS of the detected hexagons.
Column 7: Signal-to-noise ratio distribution of the detected hexagons.
Column 8: Number of detected hexagons.
}
\end{deluxetable*}

We followed the same procedure from imaging to moment map creation as of \citet{Saito22}. We applied for observatory-delivered calibration with minor manual data flagging using the appropriate {\tt CASA} version. Subsequently, we reconstructed the images using the PHANGS-ALMA imaging pipeline \citep{Leroy21a}, which includes several processes. These processes are regridding and concatenating each line visibility data, imaging with single-scale H\"{o}gbom {\tt clean} (\citealt{Hogbom74}), primary beam correction, matching spatial resolutions, unit conversion to Kelvin, cube masking, and moment map creation. We note that we did not attempt any self-calibration. After pipeline imaging, we examined all original data cubes, moment maps, and those maps convolved to a common resolution of 150~pc, basically following the quality assurance strategy described in \citet{Leroy21a}.

Before the pipeline processing, we subtracted the continuum emission in the $uv$ plane. First, we created a dirty map of each spectral window (SPW) for all Band 3 projects, which contains line and continuum emission (a total of 51 SPWs). In this step, the spectral resolution was set to 40~\kms. Subsequently, we searched all channels, and identified lines detected by the eye. After excluding channels that were $\pm$300~\kms\ near the observed frequencies of the identified lines (systemic velocity was assumed to be 1116~\kms; \citealt{Garcia-Burillo14}), we subtracted the continuum in the $uv$ plane by fitting a first-degree polynomial function using the {\tt CASA} task {\tt uvcontsub} based on the sideband of each project. The outflow velocity measured by the \coone\ and \cishort\ maps is less than $\pm$300~\kms\ \citep{Saito22}. Note that one sideband typically contains two SPWs. Consequently, we created 26 continuum-subtracted visibility datasets, which were further processed using the imaging pipeline.

The resultant spatial resolution was better than $\sim$2\arcsec\ ($\sim$144~pc). The spectral resolution of each molecular line datacube was determined as an integral multiplication of the original channel width and set to be better than 10~\kms\ depending on the achievable signal-to-noise ratios. For some fainter lines, such as H$^{13}$CN, we set $<$20~\kms. For each datacube, we measured the RMS at each channel and then measured the median value, which represents the line sensitivity. The synthesized beam size, channel width, and median noise RMS of the produced datacubes are summarized in Table~\ref{tab:data}. Here, we only list the lines used for PCA (see Section~\ref{sec:pca}), and the complete list of detected lines can be obtained from Nakajima et al. (in preparation). Note that we only used integrated intensity maps for the analysis described below. Therefore, the spectral resolution choice does not affect the PCA results presented in this \paper.

We retrieved the calibrated Multi Unit Spectroscopic Explorer (MUSE/VLT) \siii$\lambda\lambda$9069,9532/\sii$\lambda\lambda$6717,6731 ratio map\footnote{This is available from the Strasbourg astronomical Data Center (CDS); see \url{http://cdsarc.u-strasbg.fr/viz-bin/qcat?J/A+A/622/A146}.}, which traces well the structure of a 1-kpc-scale biconical ionized gas outflow \citep{Mingozzi19}.

\begin{figure*}[t!]
\begin{center}
\includegraphics[width=15cm]{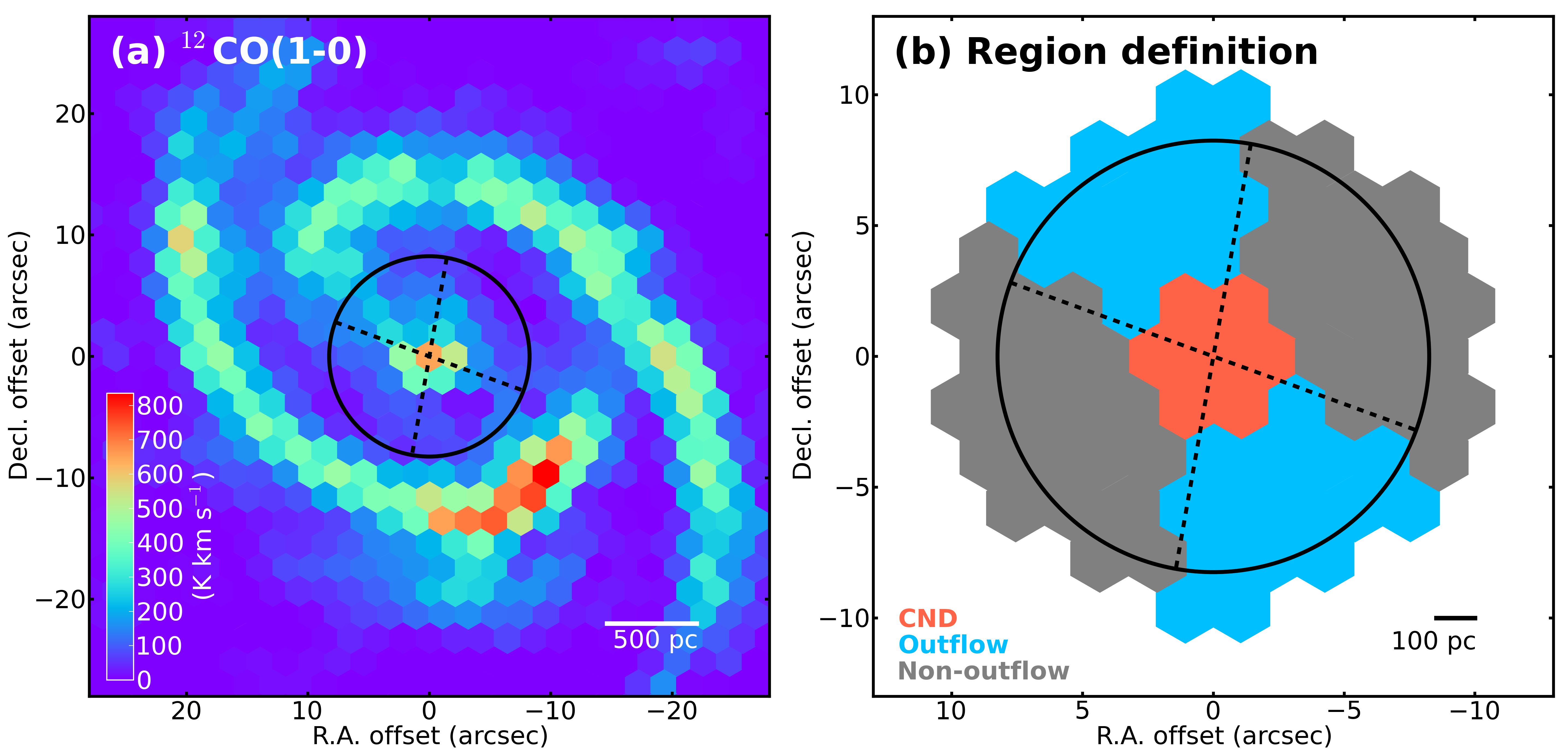}
\end{center}
\caption{
(a) \coone\ hexagonally sampled integrated intensity map of the main molecular gas disk of NGC~1068. The two dashed lines crossing the AGN position \citep{Roy98} denote the approximate outer edges of the ionized gas cones \citep{Mingozzi19}. The center position of this image is ($\alpha$, $\delta$)$_{\rm J2000}$ = (2h42m40.7132s, -0d00m47.655s). The black circle represents the field of view of the \cishort\ data (i.e., nearly central 1-kpc).
(b) Definition of subregions within the central 1-kpc of NGC~1068.
\label{fig:overall}}
\end{figure*}

\subsection{PCA} \label{sec:pca}
We convolved all integrated intensity maps and ancillary maps to a 150-pc beam  ($=$ 2\farcs08) and subsequently regridded them to a 150-pc-scale hexagonal grid \citep[following, e.g.,][]{denBrok21}, to minimize pixel correlation. When we create the integrated intensity maps, we clipped the datacubes at a signal-to-noise ratio of 4. The 150-pc resolution corresponds to ``cloud-scale" \citep[e.g.,][]{Leroy21b}. We directly regridded the \siii/\sii\ ratio map without convolution. An example of the hexagonal maps is shown in Figure~\ref{fig:overall}a. Subsequently, we extracted the central 20\arcsec\ diameter of NGC~1068 centered at the AGN position \citep{Roy98}. The extracted part is shown in Figure~\ref{fig:overall}b. This map is colorized to divide the central part into three sub-regions: circumnuclear disk (``CND"), biconical outflow (``Outflow"), and the remaining part (``Non-outflow"). The outer radius of the CND is $\sim$200~pc \citep[e.g.,][]{Garcia-Burillo14}; therefore, the central seven hexagons ($r$ $\sim$ 225~pc) cover the entire part of the CND. We used the outline of the biconical outflow defined by \citet{Das06}, which describes well the projected geometry of the cold gas outflow \citep[e.g.,][]{Saito22}. This sub-region definition is used in the discussion section (Section~\ref{sec:discussion}) but not used for feature extraction using PCA.

\begin{figure*}[t!]
\begin{center}
\includegraphics[width=17.5cm]{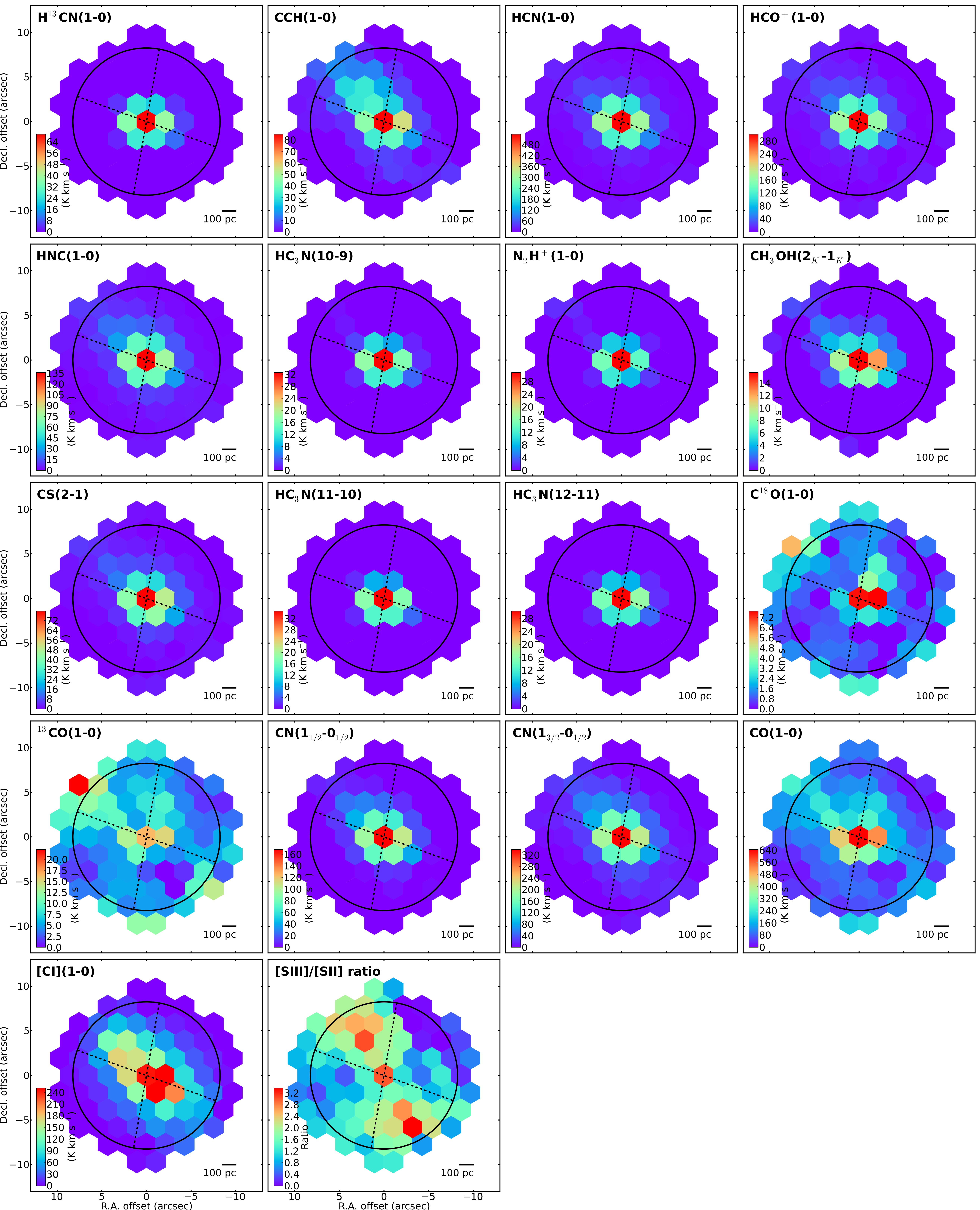}
\end{center}
\caption{
Image gallery of 16 well-detected 3-mm molecular lines in the central 1-kpc of NGC~1068 (top 16 rows) and two ancillary maps (\cishort\ and \siii/\sii\ ratio), which are known to trace the biconical outflow structure. In the top-left corner of each panel, name of the atomic and molecular line and number of detected pixels are presented. Two dashed lines crossing the AGN position \citep{Roy98} denote approximate outer edges of the ionized gas cones \citep{Mingozzi19}. The black circle represents the field of view of the \cishort\ data (i.e., nearly the central 1-kpc).
\label{fig:gallery}}
\end{figure*}

Finally, maps with fewer than twenty detected hexagons were removed, resulting in 16 molecular lines: \htcn, \cch, \hcn, \hcop, \hnc, \hcccnten, \nthp, \methanol, \cs, \hcccneleven, \hcccntwelve, \ceoone, \tcoone, \cnl, \cnh, and \coone\ (see Figure~\ref{fig:gallery}). The above three steps significantly reduce the number of dimensions and features of the input datasets, and thus, allow PCA to be sensitive to putative faint features around the center of NGC~1068-like cold gas outflows.

\begin{figure*}[t!]
\begin{center}
\includegraphics[width=18cm]{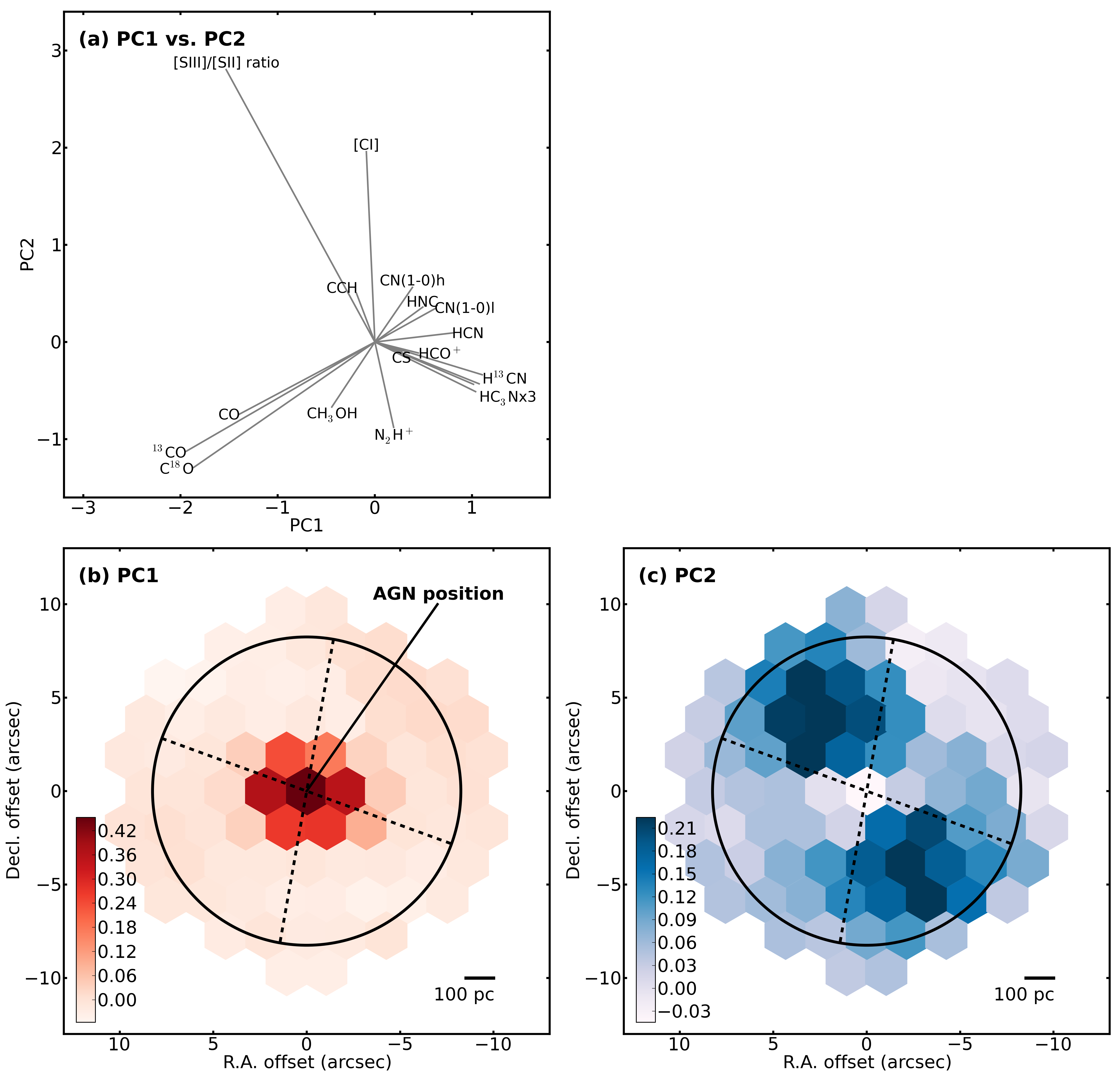}
\end{center}
\caption{
(a) The scatter plot of first two PCs of each image. Each PC is extracted by applying PCA to 16 molecular line maps and two ancillary maps (\siii/\sii\ ratio and \cishort\ intensity). PC maps are PC1 (b) and PC2 (c). Two dashed lines crossing AGN position \citep{Roy98} denote approximate outer edges of ionized gas cones \citep{Mingozzi19}. Circle represents approximate field of view of Band~8 \cishort\ map \citep{Saito22}, which has the smallest FoV among the maps described in this \paper.
\label{fig:pcamom0_map}}
\end{figure*}

Because PCA is not scale-invariant, we standardized all maps before applying PCA, i.e., all input maps had a mean value of zero and a standard deviation of unity. All nondetected hexagons were zero-padded, to match the dimensions. Subsequently, we calculated the covariance matrix and sorted its eigenvectors (or principal component using eigenvalues). The PC values were labeled as PC1, PC2, ..., starting with the largest value. Finally, we evaluated the cumulative contribution rate (CCR) of each PC to determine the amount of variance that the PCs explain. We found that the first two PCs accounted for the majority (90.0\%) of the variability in the input maps. Thus, in this study, we considered that PC1 and PC2 can adequately extract gaseous features in the central 1-kpc of NGC~1068 at 150 pc resolution. In Table~\ref{tab:pca}, we list the CCRs up to PC5 and all PC vectors.

%%%%%%%%%%%%%%%%%%%%%%%%%%%%%%
%%%%%%%%%% Results %%%%%%%%%%%
%%%%%%%%%%%%%%%%%%%%%%%%%%%%%%
\section{Results} \label{sec:results}
Figure~\ref{fig:pcamom0_map} shows the scatter plot of the first two PC values of each image as well as the feature images of the first two eigenvectors (hereafter PC maps). These data are extracted by applying PCA to 16 integrated intensity maps plus 2 ancillary maps. PC1 is in a direction that maximizes the variance of the projected data, and PC2 has a direction orthogonal to PC1 that maximizes the variance among all directions orthogonal to PC1. Thus, to understand the results of PCA, it is helpful to focus on the observables that show large absolute PC values.

\begin{deluxetable*}{lcccccc}
%\tablenum{1}
\tablecaption{cumulative contribution rates and all PC vectors up to PC5 ordered by PC2 values \label{tab:pca}}
\tablewidth{0pt}
\tablehead{
\colhead{} & \colhead{PC1} & \colhead{PC2} & \colhead{PC3} & \colhead{PC4} & \colhead{PC5}
}
%\decimalcolnumbers
\startdata
CCR               & \p0.69 & \p0.90 & \p0.97 & \p0.98 & \p0.99 \\
\siii/\sii\ ratio & \m1.53 & \p2.80 & \p2.50 & \m0.23 & \p0.08 \\
\ci\              & \m0.09 & \p1.96 & \m2.55 & \m2.00 & \m0.28 \\
\cnh\             & \p0.39 & \p0.56 & \m0.50 & \p0.89 & \p0.86 \\
\cch\             & \m0.19 & \p0.50 & \m0.62 & \p2.44 & \m2.74 \\
\hnc\             & \p0.49 & \p0.36 & \m0.25 & \p0.78 & \p0.89 \\
\cnl\             & \p0.62 & \p0.35 & \m0.75 & \p0.90 & \p0.41 \\
\hcn\             & \p0.81 & \p0.09 & \m0.04 & \p0.48 & \p1.34 \\
\cs\              & \p0.28 & \m0.10 & \m0.40 & \m0.36 & \p0.06 \\
\hcop\            & \p0.45 & \m0.11 & \p0.10 & \p1.25 & \p1.14 \\
\htcn\            & \p1.10 & \m0.34 & \p0.61 & \m0.65 & \p1.06 \\
\hcccnten\        & \p1.01 & \m0.43 & \p0.61 & \m0.47 & \m0.40 \\
\hcccntwelve\     & \p1.07 & \m0.43 & \p0.69 & \m1.07 & \m0.55 \\
\hcccneleven\     & \p1.04 & \m0.51 & \p0.90 & \m1.27 & \m1.85 \\
\methanol\        & \m0.44 & \m0.67 & \m0.84 & \m0.32 & \m0.50 \\
\coone\           & \m1.40 & \m0.74 & \m0.30 & \m0.05 & \p0.20 \\
\nthp\            & \p0.19 & \m0.88 & \p0.92 & \p0.33 & \m0.36 \\
\tcoone\          & \m1.94 & \m1.12 & \m0.06 & \m0.15 & \p0.35 \\
\ceoone\          & \m1.86 & \m1.29 & \m0.11 & \m0.49 & \p0.28 \\
\enddata
%\tablecomments{CCR shows each PC's contribution to the total variance.}
\end{deluxetable*}

\subsection{PC1 represents CND}
The most deviated observables in the PC1 direction are \hcccn\ and \htcn\ (positive deviation) and CO isotopologues and \siii/\sii\ ratio (negative deviation). The \hcccn\ lines are known to be centrally concentrated around the CND at this spatial scale \citep[e.g.,][]{Takano14}. However, CO isotopes \citep[e.g.,][]{Tosaki17} and the \siii/\sii\ ratio \citep{Mingozzi19} are known to be spatially extended or not clearly concentrated. These observed trends reported in the literature are consistent with the visual tendencies of the PC1 map. Therefore, PCA extracts the CND itself as the most prominent gaseous feature in the central 1-kpc of NGC~1068. This explanation is supported by the fact that other high-dipole molecules that are also known to coincide with the CND, such as \hcn\ and \hcop \citep{Kohno08}, show high PC1 values.

The \cishort\ map shows an approximately zero PC1, potentially because of the presence of bright extended features along the biconical outflow that are extended differently than CO \citep{Saito22}. Interestingly, a dense gas tracer, \nthp\ \citep[e.g.,][]{Pety17,Kauffmann17,Barnes20,Tafalla21}, does not show a high PC1, different from other well-known dense gas tracers.

%Figure~\ref{fig:pcaratio_podium} shows the three highest PC1 maps---\htcn, \hcccn\ lines, and \hcn---and the lowest map, \tcoone. As described above, the highest PC1 maps are characterized well by a central concentration. The three highest PC1 maps and the fourth map (\hcop) divided by the \coone\ map are shown in the top panels of Figure~\ref{fig:pcaratio_podium}. We find the same concentrations in these ratio maps (with some possible elongation to the bicone axis in the \hcn/\coone\ map).

\begin{figure*}[t!]
\begin{center}
\includegraphics[width=18cm]{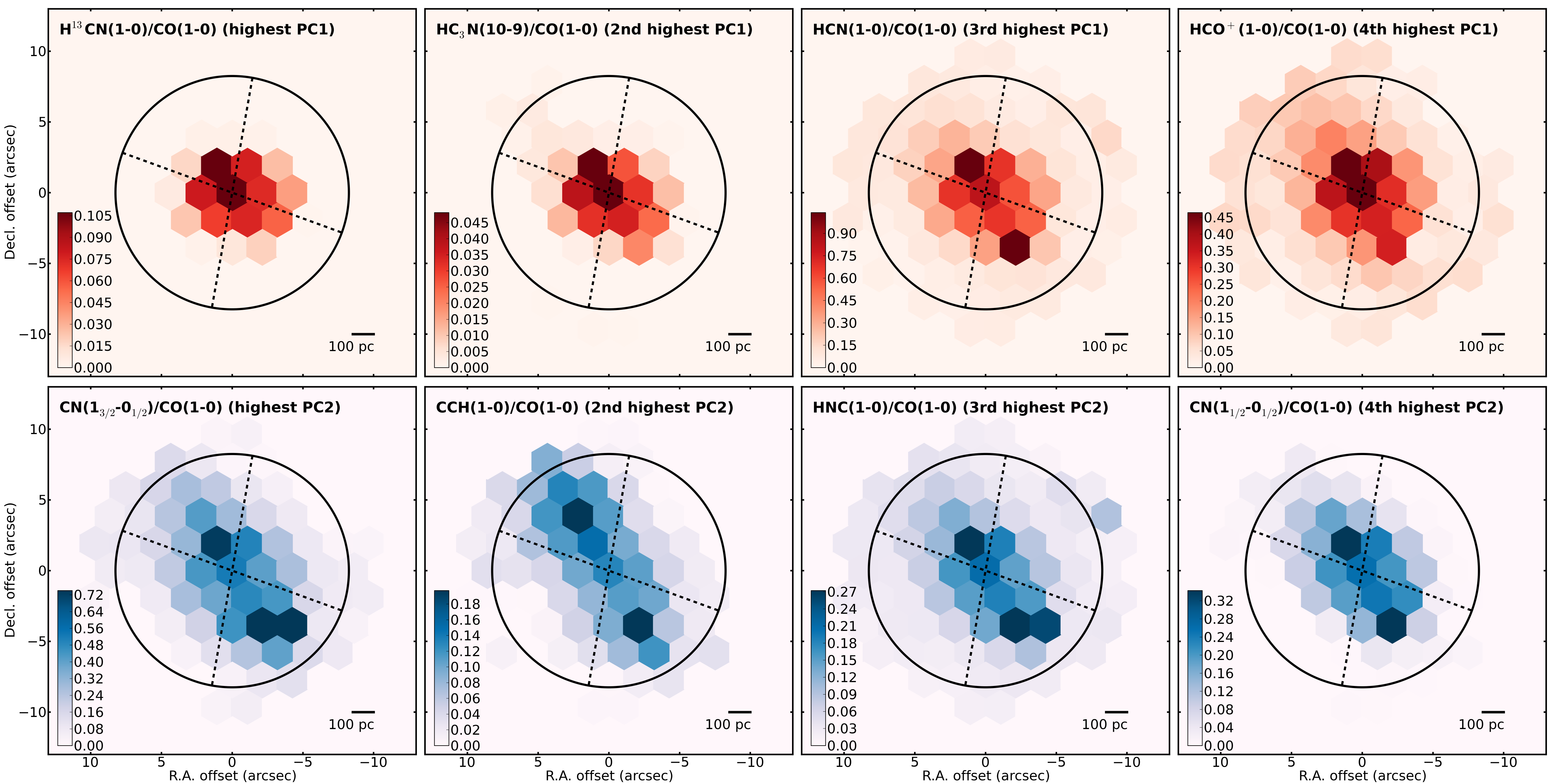}
\end{center}
\caption{	
(top) Four highest and most lowest PC1 maps divided by \coone\ map as determined by PCA. (bottom) Corresponding PC2 maps.
\label{fig:pcaratio_podium}}
\end{figure*}

The molecular line maps that have the highest PC values---\htcn, \hcccn\ lines, \hcn, and \hcop--- are characterized well by a central concentration. In the top panels of Figure~\ref{fig:pcaratio_podium}, we show four molecular line maps, that have the highest PC1 values, divided by the \coone\ map, and corresponding PC2 maps are shown in the bottom panel. Here \hcccnten\ represents the other \hcccn\ maps as these have similar PC1 and PC2 values. The line ratio maps still show the central concentration in comparison with the corresponding line ratio maps that have the highest PC2 values.

\subsection{PC2 represents outflow}
As noticeable in Figure~\ref{fig:pcamom0_map}(a), the second-most prominent feature extracted by PCA is highly positively correlated with the \siii/\sii\ ratio map. This map displays a metallicity-independent ionization parameter distribution \citep{Mingozzi19}. In the case of the NGC~1068 center, it probably traces the biconical ionized gas outflow. The second-most deviated observable in the PC2 direction is the \cishort\ map. \citet{Saito22} have shown that \cishort\ gas traces the biconical feature along the outflow, and thus PC2 probably extracts a biconical feature along the outflow direction. This is consistent with the visual tendency of the PC2 map (Figure 3c), which shows two symmetric peaks a few hundred parsecs away from the AGN position.

The \cnh, \cch, \hnc, and \cnl molecular line maps (Figure~\ref{fig:gallery}) and their line ratio maps (Figure~\ref{fig:pcamom0_map}) are similar to the PC2 map. Interestingly, all radicals used in PCA are selected. The remaining molecule, HNC, is also a unique molecule among the input molecules because it is an isomer of HCN. In Section~\ref{sec:discussion}, we discuss plausible explanations for the enhancement of these molecules in the AGN outflow of NGC~1068, as well as for the lower PC2 maps, i.e., CO isotopes, \methanol, and \nthp.

We note that a bright spot in the north-eastern pixel of the \tcoone, \ceoone, and \coone\ (Figure~\ref{fig:gallery}) is probably a part of the northern bar-end (Figure~\ref{fig:overall}a). This spot is not captured in the PC1 and PC2 maps, and thus this is not the primary reason why these molecules systematically show lower PC1 and PC2 values. Apart from this, the bar itself may affect the PCA with 150~pc resolution maps (see the molecular bar presented in \citealt{Garcia-Burillo14}), although it is reported that the \cishort\ and \siii/\sii\ ratio maps mainly trace the outflow, not the bar \citep{Mingozzi19,Saito22}. Differentiating the bar from the outflow should be done with higher angular resolution data. In this \paper\ we regard that the PC2 map represents the outflow.

%\subsection{CS -- a molecule uncorrelated to PC1 and PC2}
%\comment{TBE.}

%%%%%%%%%%%%%%%%%%%%%%%%%%%%%%
%%%%%%%%% Discussion %%%%%%%%%
%%%%%%%%%%%%%%%%%%%%%%%%%%%%%%
\section{Discussion} \label{sec:discussion}
Here, we discuss a possible mechanism that can explain all above-mentioned trends.

Based on observational studies in the literature, a kpc-scale biconical structure driven by a radio jet \citep[e.g.,][]{Gallimore1996b,Gallimore1996a} is considered to penetrate the disk plane \citep[e.g.,][]{Schinnerer00,Das06,Garcia-Burillo14,Garcia-Burillo19}. The biconical structure is likely driven by a jet-driven expanding bubble which heats the surrounding ISM and creates shocks resulting in multiphase outflows (see also numerical simulations done by \citealt{Mukherjee16}). By its strong ultraviolet (UV) and X-ray radiation fields as well as large-scale shocks \citep[e.g.,][]{Barbosa14,May17}, this outflowing bicone naturally causes photo- and shock-dissociation \citep[e.g.,][]{Garcia-Burillo17,Saito22} and ionization \citep[e.g.,][]{Mingozzi19} of molecular clouds within the disk. We suggest that the mechanism of molecular enhancement/suppression can be explained by dissociative processes occurring in the biconical outflow. Here, we discuss individual molecules in this context.

\subsection{Molecules enhanced in outflow}
\begin{figure*}[t!]
\begin{center}
\includegraphics[width=18cm]{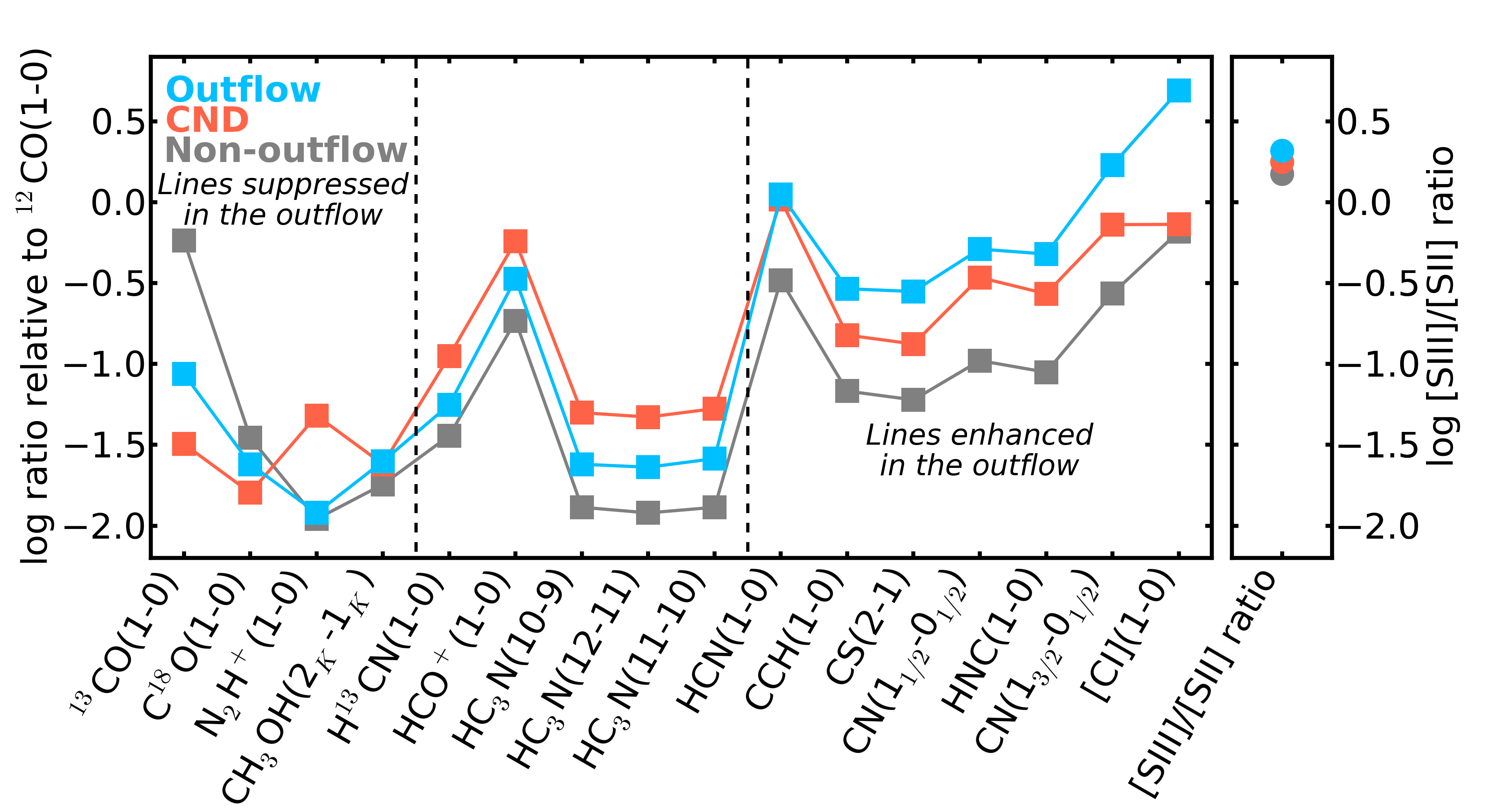}
\end{center}
\caption{
Integrated line intensities relative to \coone\ integrated intensity in each subregion within the central 1-kpc of NGC~1068. Here, we plot the maximum ratio within each subregion. Lines are ordered by differences between ``Outflow" (blue) and ``Non-outflow" (grey), i.e., lines on the right-hand (left-hand) side are enhanced (suppressed) in outflow. In the right panel, we plot the \siii/\sii\ ratio for reference. This order is similar to that by PC2 values, supporting that emission lines positively correlated to PC2 map are enhanced in the outflow.
\label{fig:line_graph}}
\end{figure*}

{\it C$_2$H enhancement:} Enhancement in \cch\ in the outflow of NGC~1068 was previously reported in \citet{Garcia-Burillo17} using high-angular resolution ALMA datasets. Based on chemical modeling, the authors concluded that the observed high abundance of C$_2$H can only be achieved in a dense molecular gas heavily irradiated by UV/X-ray photons and/or in icy mantle sputtering due to shocks driven by the jet--ISM interaction. These processes dissociate CO and release C atoms and C$^+$ ions, which are used to form C$_2$H \citep[e.g.,][]{Fuente93}. This CO dissociation scenario is preferable for explaining the observed high \cch/\coone\ ratio distribution (Figure~\ref{fig:pcaratio_podium}) and the observed high \cishort/CO ratio distribution within the outflow of NGC~1068 \citep{Saito22}. Figure~\ref{fig:line_graph} shows the maximum value of the integrated line intensity divided by the \coone\ line integrated intensity in the three specific regions shown in Figure~\ref{fig:overall}b. In the ``Outflow", the \cch/\coone\ ratio is $\sim$0.6 dex higher than in the ``Non-outflow". In the case of the \cishort/CO ratio, the difference is almost an order of magnitude. Both trends quantitatively support the CO dissociation scenario, which results in bright \cch\ and \ci\ lines in  the outflow.

According to the regions defined in Figure~\ref{fig:overall}b, the lines in Figure~\ref{fig:line_graph} are ordered by the maximum ratio in the ``Outflow" divided by the maximum ratio in the ``Non-outflow". Thus, the lines on the right-hand (left-hand) side tend to be enhanced (suppressed) in the ``Outflow" region. This order is similar to that by the PC2 values (see Table~\ref{tab:pca}), and thus, supports the results of PCA in a complementary manner.

{\it CN enhancement:} A similar explanation to \cch\ above (except for the dust sputtering) can be used for the CN enhancement. CN is considered to primarily form from the photodissociation of HCN and neutral--neutral reactions with some molecules \citep[e.g.,][]{Aalto02}. The production efficiency of CN is highly dependent on the number of free C atoms and C$^+$ ions in the case of neutral--neutral reactions such as CH $+$ N $\rightarrow$ CN $+$ H, which is considered as one of the most important reaction pathways to CN in diffuse PDR, i.e., irradiated ISM \citep[e.g.,][]{Fuente93,Boger05}. Thus, CN abundance can be enhanced in a molecular gas irradiated by UV/X-ray radiation. Based on PCA and Figure~\ref{fig:line_graph}, both CN hyperfine transitions are enhanced in the ``Outflow" region compared to the HCN line, which supports the dissociation-enhanced CN scenario in the cold gas outflow. A similar CN enhancement is reported in the cold gas outflow of the nearest ULIRG QSO Mrk~231 \citep{Cicone20}.

{\it HNC enhancement:} Based on PCA, the HNC map is more positively correlated with the PC2 map than the HCN map, although the statistical difference is small. Based on Figure~\ref{fig:line_graph}, the HNC/CO difference between ``Outflow" and ``Non-outflow" is $\sim$0.8 dex, which is larger than the HCN/CO difference of $\sim$0.5 dex. Thus, both analyses suggest that HNC is more enhanced in the outflow than HCN.

The HNC/HCN ratio is observationally known to correlate with the gas kinetic temperature (decreasing ratio with increasing gas temperature due to different energy levels and an isomerization barrier; e.g., \citealt{Hacar20}). However, in irradiated ISM, free electrons play a crucial role in enhancing the HNC/HCN ratio to approximately unity by the dissociative recombination (DR) of HCNH$^+$ \citep[e.g.,][]{Herbst00}. The same scenario was proposed by \citet{Aalto02} to explain the observed high HNC/HCN intensity ratio for nearby luminous infrared galaxies. As the DR can produce HNC and HCN with almost equal branching ratios \citep[e.g.,][]{Mendes12}, the observed abundance ratio also becomes almost unity. Thus, we suggest that the high HNC/HCN ratios within the bicone of NGC~1068 are driven by the DR of HCNH$^+$, i.e., dissociation enhancement of HNC. Although this scenario should be further investigated using higher quality observational data, this fits well with the explanations for the other enhanced molecules, discussed above. Note that CN can also be formed in the DR of HCNH$^+$ \citep[e.g.,][]{Herbst95}, which is another possible explanation to the enhancement of CN in the ``Outflow."

\subsection{Molecules suppressed in outflow}
{\it CO suppression:} The gas-phase abundance of CO and its isotopologues is controlled by photodissociation caused by UV photons \citep[e.g.,][]{Visser09} and shock dissociation \citep[e.g.,][]{Hollenbach80}. The destruction of CO and the production of C atoms are revealed in the outflow of NGC~1068 \citep{Saito22} and other galaxies \citep[e.g.,][]{Cicone18,Izumi20}. Considering isotope-selective photodissociation (i.e., less abundant molecules are less self-shielded; e.g., \citealt{Bally82}), the CO isotopologues are also expected to be destroyed even with higher dissociation rates. The dissociation-suppressed scenario naturally suggests a strong anti-correlation between the CO maps and the PC2 map (see also Table~\ref{tab:pca}).

{\it CH$_3$OH suppression:} Gas-phase CH$_3$OH becomes abundant when interstellar icy mantles rich in CH$_3$OH are sublimed by radiation or shock heating \citep[e.g.,][]{Viti11}, whereas the suppression of CH$_3$OH is expected in clouds irradiated by a strong UV radiation field without shielding ($A_v$ $\sim$ 5; \citealt{Martin09}). In the case of NGC~1068's center, the CH$_3$OH map anti-correlates with the PC2 map, i.e., CH$_3$OH is suppressed in the outflow. However, this does not necessarily indicate the absence of shocks, as shock-enhanced CH$_3$OH can be rapidly destroyed by UV radiation. Radiation heating and shock heating are expected to coexist in jet-driven multi-phase outflows, as shown by different molecular and atomic line observations \citep{Garcia-Burillo17,Saito22} and numerical simulations of jet--ISM interaction phenomena \citep[e.g.,][]{Mukherjee16}. Thus, we suggest that shock CH$_3$OH enhancement may occur in the outflow. However, dissociation suppression dominates.

Dissociative high-speed shocks might be expected to destroy molecules, leading to the suppression of CH$_3$OH \citep[e.g.,][]{Hollenbach80}. However, in this case, most molecules other than CH$_3$OH are also destroyed, which is inconsistent with the molecular-rich nature of the NGC~1068 outflow. Thus, we suggest that ``non-dissociative shocks and dissociative UV/X-ray radiation" best probably describe the chemistry of the cold gas outflow of NGC~1068. This is consistent with conclusions based on high-spatial resolution \cch\ observations \citep{Garcia-Burillo17}.

{\it  N$_2$H$^+$ suppression:} Similar to the DR of HCNH$^+$, its reaction with free electrons dissociates N$_2$H$^+$. Thus, if free elections play an important role in enhancing HNC, the dissociation suppression of N$_2$H$^+$ (and another molecular ion HCO$^+$) is plausible. A large number of free electrons in the outflow are expected, based on the high \siii/\sii\ ratio along the bicone \citep{Mingozzi19}. However, reactions with neutral molecules and molecular ions and the effect of cosmic rays are also important for understanding the chemistry of these molecular ions, not only DR. Thus, accurate column density measurements \citep[e.g.,][]{Nakajima18} and complex chemical modeling (e.g., as done by \citealt{Harada21}) are required to further constrain the behavior of molecular ions in the cold gas outflow of NGC~1068.

\subsection{Molecules enhanced/suppressed in CND}
As briefly described above, molecules concentrated to the CND are high-dipole molecules. Molecules negatively correlated are CO and its isotopologues which are easily destroyed by UV photons and X-rays. These trends support the findings reported in \citet{Takano14}, i.e., the CND contains regions where \hcccn\ molecules are effectively shielded from the strong AGN radiation. The different correlation strength among the input maps may imply the chemistry and physics around the central AGN \citep[e.g., X-ray dominated chemistry, shock chemistry, and high-temperature chemistry;][]{Meijerink05,Harada13,Izumi13,Huang22}. For the bright central region of NGC~1068, constraining the excitation of these molecules with higher angular resolution data is an interesting future direction.

\citet{Huang22} recently reported that the molecular gas outflow in the CND is shock-driven. Thus, interestingly, the putative jet-ISM interaction between the radio jet and the CND naturally explains their findings within the CND and our findings outside the CND consistently. This further supports the results of the PCA and our interpretation described in this \paper.

%%%%%%%%%%%%%%%%%%%%%%%%%%%%%%
%%%%%%%%%% Summary %%%%%%%%%%%
%%%%%%%%%%%%%%%%%%%%%%%%%%%%%%
\section{Summary} \label{sec:summary}
In this paper, we present our 150-pc-resolution ALMA Band~3 spectral scans for the central kpc of the nearby type-2 Seyfert galaxy, NGC~1068, supplemented with archival Band~3 datasets. To characterize lines related to the cold gas outflow of this galaxy, we employ the PCA algorithm and successfully categorized the detected lines into (1) lines related to the CND and (2) lines related to the outflow.

We find that the lines enhanced in the CND are mostly high-critical density tracers (e.g., H$^{13}$CN, HC$_3$N, and HCN), whereas those in the outflow are CN, C$_2$H, HNC, and possibly HCN. We also find that CO and its isotopologues, CH$_3$OH, and N$_2$H$^+$ are suppressed in the outflow.

We suggest that all enhanced and suppressed molecules currently detected in the cold gas outflow can be explained by a coherent scenario: non-dissociative shocks and dissociative UV/X-ray radiation driven by the jet--ISM interaction. Thus, we provide observational evidence of how negative AGN feedback occurring in the central 1-kpc of a galaxy chemically changes the molecular ISM of the host galaxy. However, our discussions rely on fluxes and flux ratios, and thus, we need accurate column density measurements and complex chemical modeling to provide quantitative constraints on this scenario.

To emphasize the large-scale structure in the central 1-kpc, we degrade the original molecular line intensity maps. This allows PCA to find a faint extended structure along the well-known biconical outflow from the input maps. However, higher angular resolution and higher sensitivity maps are required to accurately understand the effects of the negative AGN feedback, i.e., the changes in the molecular clouds affected by the AGN outflow.

Finally, we note that the results of PCA described in this paper change, when the sensitivity and angular resolution of the input data improve. Especially, molecules with less number of detected hexagons are more affected. The explanations for these molecules should be read with caution.

%%%%%%%%%%%%%%%%%%%%%%%%%%%%%%
%%%%%% Acknowledgments %%%%%%%
%%%%%%%%%%%%%%%%%%%%%%%%%%%%%%
\begin{acknowledgments}
This work was supported by NAOJ ALMA Scientific Research Grant Numbers 2021-18A. NH acknowledges support from JSPS KAKENHI Grant Number JP21K03634. KK acknowledges support from JSPS KAKENHI Grant Number JP17H06130. The work of YN was supported by NAOJ ALMA Scientific Research Grant Numbers 2017-06B and JSPS KAKENHI grant Number JP18K13577. The work of KB was supported in part by the JSPS KAKENHI Grant Number JP21K03547. TT acknowledges support from JSPS KAKENHI Grant Number JP20H00172 and NAOJ ALMA Scientific Research Grant Numbers 2020-15A.
This work was supported by NAOJ ALMA Scientific Research Grant Numbers 2021-18A. We would like to thank Editage (www.editage.com) for English language editing. This study was supported by NAOJ ALMA Scientific Research Grant Number 2021-18A. This paper makes use of the following ALMA data: ADS/JAO.ALMA\#2011.0.00061.S,\\
ADS/JAO.ALMA\#2012.1.00657.S,\\
ADS/JAO.ALMA\#2013.1.00060.S,\\
ADS/JAO.ALMA\#2013.1.00279.S,\\
ADS/JAO.ALMA\#2015.1.00960.S,\\
ADS/JAO.ALMA\#2017.1.00586.S,\\
ADS/JAO.ALMA\#2018.1.01506.S,\\
ADS/JAO.ALMA\#2018.1.01684.S, and\\
ADS/JAO.ALMA\#2019.1.00130.S.
ALMA is a partnership of ESO (representing its member states), NSF (USA) and NINS (Japan), together with NRC (Canada), MOST and ASIAA (Taiwan), and KASI (Republic of Korea), in cooperation with the Republic of Chile. The Joint ALMA Observatory is operated by ESO, AUI/NRAO and NAOJ. This research has made use of the NASA/IPAC Extragalactic Database (NED), which is funded by the National Aeronautics and Space Administration and operated by the California Institute of Technology. The National Radio Astronomy Observatory is a facility of the National Science Foundation operated under cooperative agreement by Associated Universities, Inc. This research has made use of the SIMBAD database, operated at CDS, Strasbourg, France. Data analysis was in part carried out on the Multi-wavelength Data Analysis System operated by the Astronomy Data Center (ADC), National Astronomical Observatory of Japan.
\end{acknowledgments}

\vspace{5mm}
%\facilities{ALMA, VLA, VLT(MUSE)}

\software{
{\tt ALMA Calibration Pipeline},
{\tt Astropy} \citep{Astropy13,Astropy18},
{\tt CASA} \citep{McMullin07},
{\tt NumPy} \citep{Harris20},
{\tt PHANGS-ALMA Pipeline} \citep{Leroy21a},
%{\tt PHANGS-ALMA Total Power Pipeline} \citep{Herrera20},
{\tt SciPy} \citep{Virtanen20},
{\tt spectral-cube} \citep{Ginsburg19},
{\tt radio-beam}
}

%%%%%%%%%%%%%%%%%%%%%%%%%%%%%%
%%%%%%%%%% Appendix %%%%%%%%%%
%%%%%%%%%%%%%%%%%%%%%%%%%%%%%%
\appendix
\section{Other minor feature maps extracted by PCA} \label{appendix:all_pc}
\begin{figure*}[t!]
\begin{center}
\includegraphics[width=18cm]{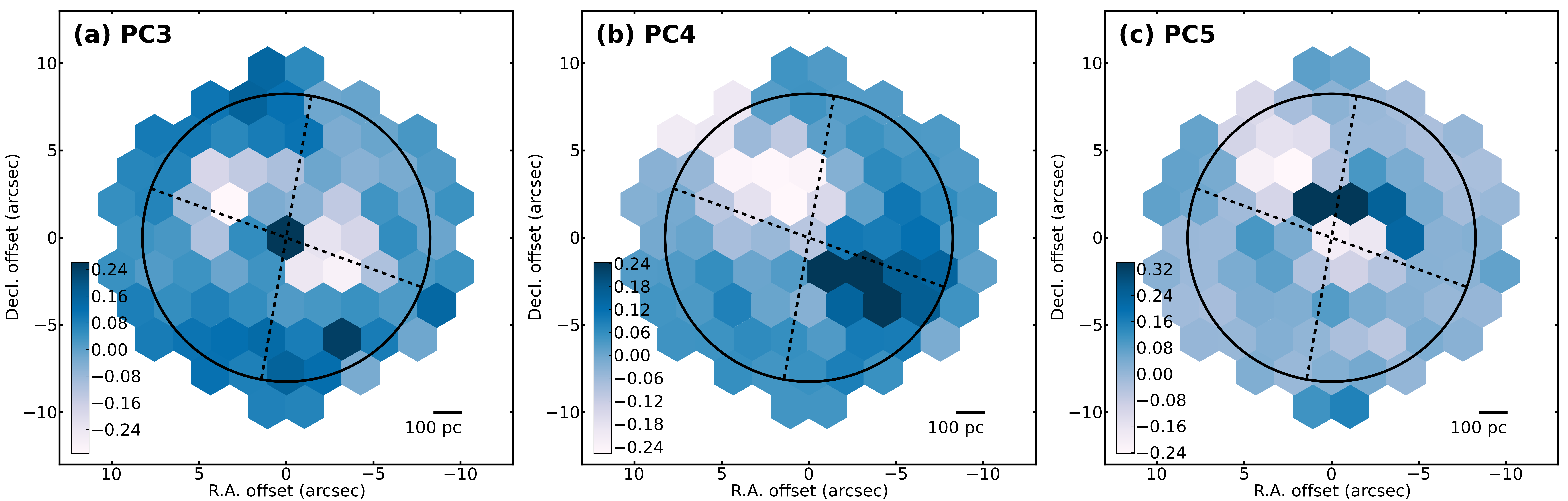}
\end{center}
\caption{
Minor PC maps; PC3 (a), PC4 (b), and PC5 (c). Two dashed lines crossing AGN position \citep{Roy98} denote approximate outer edges of ionized gas cones \citep{Mingozzi19}. Circle represents approximate field of view of Band~8 \cishort\ map \citep{Saito22}, which is smallest among maps described in this \paper.
\label{fig:appendix_pcs}}
\end{figure*}

Here we describe three PC maps that account for the minor fraction (10.0\%) of the variability in the input 18 maps (Figure~\ref{fig:appendix_pcs}). PC3 probably captures a peak at the AGN position. However, the explanation is not straightforward because extended emission also show positive values. PC4 shows a positive blob in the south and a negative blob in the north. Interestingly, two maps showing outflow-like features (\cishort\ and \cch) are classified as the most positively and negatively correlated maps. This may imply the chemical difference in the northern and southern part of the outflow, although note that PC4 and PC5 only account for 2\% of the variability in the input maps. Higher angular resolution maps are useful to uncover these possible features.

%%%%%%%%%%%%%%%%%%%%%%%%%%%%%%
%%%%%%%% bibliography %%%%%%%%
%%%%%%%%%%%%%%%%%%%%%%%%%%%%%%
\bibliography{n1068_pca}{}
\bibliographystyle{aasjournal}

\end{document}